\documentstyle[12pt,emulateapj]{article}

\newcommand{\gtilde}
 {~ \raisebox{-1ex}{$\stackrel{\textstyle >}{\sim}$} ~}
\newcommand{\ltilde}
 {~ \raisebox{-1ex}{$\stackrel{\textstyle <}{\sim}$} ~}

\begin{document}

\submitted{To Appear in The Astrophysical Journal Letters}

\title{\small
TeV Burst of Gamma-Ray Bursts and Ultra High Energy Cosmic Rays}

\author{Tomonori Totani}
\affil{Department of Physics, School of Science,
The University of Tokyo, Tokyo 113-0033,
Japan \\ E-mail: totani@utaphp2.phys.s.u-tokyo.ac.jp}

\footnotesize

\begin{abstract}
Some recent experiments detecting very high energy (VHE) gamma-rays above
10--20 TeV independently reported VHE bursts for
some of bright gamma-ray bursts (GRBs). If these signals are truly from GRBs, 
these GRBs must emit a much larger amount of energy
as VHE gamma-rays than in the ordinary photon energy range 
of GRBs (keV--MeV). We show that such extreme phenomena can be reasonably 
explained by synchrotron radiation of protons accelerated to 
$\sim 10^{20-21}$ eV, which has been predicted by Totani (1998a).
Protons seem to carry about $(m_p/m_e)$ times larger energy than electrons,
and hence the total energy liberated by one GRB becomes as large as 
$\sim 10^{56} (\Delta \Omega / 4 \pi)$ ergs. Therefore a strong beaming of
GRB emission is highly likely. Extension of the VHE spectrum
beyond 20 TeV 
gives a nearly model-independent lower limit of the Lorentz factor of GRBs, as
$\gamma \gtilde 500$.
Furthermore, our model gives the correct energy range and time variability
of ordinary keV--MeV
gamma-rays of GRBs by synchrotron radiation of electrons.
Therefore the VHE bursts of GRBs strongly support the hypothesis
that ultra high energy cosmic rays observed on the Earth are produced by GRBs.
\end{abstract}


\section*{}
Two major ground-based cosmic ray/gamma-ray detectors, the Tibet air shower
array (Amenomori et al. 1996)
and the HEGRA AIROBICC Cherenkov array (Padilla et al. 1998)
have independently reported
significant excesses of 10--20 TeV gamma-rays  coincident
with some GRBs both in direction and burst time.
There are 57 GRBs detected by the BATSE (Meegan et al. 1996) in the
field of view of the Tibet array, and
the Tibet group found that several GRBs out of the 57 GRBs 
show significant excess of 10 TeV gamma-ray events 
with a time scale of $\sim$ 10 seconds. The statistical 
significance was estimated to be about 6 sigma. There are four GRBs in the 
field of view of the HEGRA array, and the HEGRA group observed
11 gamma-ray like events above 20 TeV in a 4-minutes bin coincident with
GRB920925c in the GRANAT/WATCH GRB catalog (Sazonov et al. 1998),
whereas only 0.93
events are expected as background (5.4 sigma).  The chance probability
taking into account an appropriate trial factor was estimated as 0.3 \%.
Seven out of the 11 gamma-ray events are
clustered within 22 seconds, suggesting that the burst time scale is $\sim$
10 seconds also for this GRB. Considering the fact that two different
experimental groups independently reported similar significant signals 
of 10--20 TeV gamma-rays, now we must seriously take into consideration these
very high energy (VHE) emissions. Although further observations are necessary
to confirm these indications of VHE bursts, in this letter we show that
these VHE bursts, if confirmed, would 
result in a drastic change of our picture of GRBs.

There is a serious problem in interpreting these VHE gamma-rays.
The cosmological origin of GRBs has almost been confirmed
(Metzger et al. 1997; Kulkarni et al. 1998) and most of GRBs
are considered to come from large distances beyond redshift $z \sim 1$.
It is well known that VHE gamma-rays above $\sim$ TeV from such cosmological
distances are not visible because they suffer serious attenuation by
electron-positron pair creation with the intergalactic infrared background
radiation (Stecker, de Jager, \& Salamon 1992 and references therein). 
A recent calculation (Stecker \& de Jager 1998) shows that
the optical depth $\tau$ for 10 TeV gamma-rays is about 5--10 (corresponding 
to an attenuation factor of $e^{- \tau} \sim
7\times 10^{-3}$--$5 \times 10^{-5}$) for
the source redshift of $z = 0.1$--0.3, and hence there is 
a horizon for such gamma-rays at $z \sim 0.2$.
Since the GRB920925c has a large energy fluence (Sazonov et al. 1998)
of $2.1 \times 10^{-5} \ \rm erg \ cm^{-2}$ in 20--60 keV (typical
fluence range of the BATSE catalog is $10^{-7}$--$10^{-4}\ \rm erg \ cm^{-2}$
in 50--300 keV),
it is likely that the GRBs having the VHE emission are nearby bursts within
$z \ltilde 0.2$. In fact, with typical distance scales of $z_{\max}$ = 1--3,
where $z_{\max}$ is the redshift of the most distant GRBs 
observed by the BATSE, a few percent of 
the BATSE GRBs are expected to be located around $z \sim 0.1$--0.3. 
This is consistent with
the fact that only a few out of 57 BATSE GRBs in the Tibet data 
seem to have VHE emission.
It should be noted that, although the GRBs having the VHE emission must be
the nearest bursts, the flux of keV--MeV gamma-rays is not necessarily the
brightest because there may be intrinsic luminosity dispersion of GRBs.

The energy fluence of 10 TeV gamma-rays from GRB920925c was 
estimated (Padilla et al. 1998)
as $6.9 \times 10^{-6} \ \rm erg \ cm^{-2}$, which is comparable
to that in the soft gamma-ray range of 20--60 keV ($2.1 \times 10^{-5}
\ \rm erg \ cm^{-2}$). 
The VHE fluence of a few GRBs observed by the Tibet array was estimated as
$10^{-6}$--$4\times 10^{-5}\ \rm erg \ cm^{-2}$ (Amenomori et al.
1996), which is also comparable
to that of soft gamma-rays of relatively bright GRBs. Furthermore, we must
take into account the fact that the intergalactic attenuation factor of
10--20 TeV gamma-rays is likely as small as $\sim 10^{-2}$--$10^{-3}$ 
for $z \sim$ 0.1--0.2. This suggests that
the energy emitted as VHE gamma-rays is much larger than that emitted as
keV--MeV gamma-rays by a factor of $\sim 10^2$--$10^3$. This emission is
extraordinarily energetic, but in the following we show that such VHE emission
can be reasonably explained by synchrotron radiation of $\sim 10^{20-21}$ eV
protons in GRBs. The origin of ultra high energy cosmic rays (UHECRs),
whose energy extends beyond $10^{20}$ eV, is one of the
most interesting mysteries in astrophysics as well as GRBs, 
and the observations of 10 TeV gamma-rays by the Tibet 
and HEGRA groups strongly suggest that these two mysterious phenomena have a
common origin, as it has been speculated (Waxman 1995; 
Milgrom \& Usov 1995; Vietri 1995).

Since the energy emitted as 10 TeV gamma-rays is much larger than that of
soft gamma-rays, the VHE emission is likely produced by protons while
the keV--MeV emission is produced by electrons. The current theory of
GRBs (e.g., Piran 1997)
attributes the origin of gamma-ray emission to dissipation of the
kinetic energy of ultra relativistic bulk motion with a Lorentz factor of
$\gamma \sim 10^2$--$10^3$, 
in internal shocks generated by relative difference in
speeds of ejected matter. Therefore at least in the initial stage protons
must carry much larger energy 
than electrons by a factor of $\sim m_p/m_e \sim 2,000$. 
The emission mechanism of keV--MeV gamma-rays is generally
considered to be electron synchrotron.
If the energy carried by protons is effectively emitted as VHE photons before
the equipartition between protons and electrons is achieved, 
the very strong flux of 10 TeV gamma-rays observed by the two detectors
is naturally explained. Since the energy emitted as keV--MeV gamma-rays
is typically $10^{52-53}$ erg for cosmological GRBs, the total energy
liberated by one GRB event must be as large as $\sim 10^{56}$ erg if radiation
is isotropic. This energy may seem extremely large, but if GRBs are 
strongly beamed, such a huge amount of energy can be supplied. For example, the
`microquasar model' (Paczy\'{n}ski 1998)
can produce about $5 \times 10^{54}$ erg and
the above energy can be explained if the beaming factor, 
$4\pi/\Delta \Omega$, is larger than
$\sim$ 20. Such beaming factor is not unreasonable because the origin of energy
of the microquasar model is rotation energy of a Kerr
black hole with a mass of $\sim$ 10 $M_\odot$. 
Radiation likely deviates from spherical symmetry because of
the rotation. 

In fact, we have pointed out (Totani 1998a) that the energy transfer
between protons and electrons may be a quite inefficient process and
strong afterglow of GRBs in TeV range may exist by synchrotron radiation
of protons accelerated up to $10^{20-21}$ eV. The VHE bursts beyond 10 TeV
suggested by the Tibet and HEGRA experiments
may also be explained by the proton synchrotron radiation.
However, in the previous work
we have considered the afterglow phase, i.e., the external shock
generated by collisions of ejected matter and ambient interstellar matter.
In this case, the cooling time scale of such protons is about a few days and
hence the afterglow model cannot explain the observations of the two detectors
showing a burst-like time structure of $\sim$ 10 seconds. These VHE 
emissions are coincident in time with keV--MeV GRBs
(Amenomori et al. 1996; Padilla et al. 1998), and should be
explained by internal shocks rather than external shocks.
Here we present an internal shock model which can
explain the observed VHE emission by proton synchrotron as well as
the ordinary keV--MeV emission by electron synchrotron.

Suppose that the total energy $E$ is isotropically 
emitted during a time interval $T = 10 T_{10}$ sec which is the typical
duration of GRBs, 
and this energy is converted into kinetic energy of relativistic 
bulk motion with
a Lorentz factor of $\gamma = 300 \gamma_{300}$. As discussed above,
we consider a typical case with $E = 10^{56}E_{56}$ erg. The following
analysis is not affected by the unknown beaming factor if we scale the total
energy appropriately. If the Lorentz factor
has fluctuations of $\Delta \gamma \sim \gamma$, faster ejecta will
catch up with slower ejecta at 
\begin{equation}
R \sim \gamma^2 T = 2.7 \times 10^{16}
\gamma_{300}^2 T_{10} \rm \ cm \ . 
\end{equation}
(We use the natural units of $c=\hbar=1$.)
Therefore the energy density of internal shocks
can be estimated as $\rho_{\rm lab} \sim E/(4 \pi R^2 T)$ at the laboratory 
frame. 
The restframe energy density $\rho_{\rm rest} \sim \rho_{\rm lab}
/ \gamma^2$ is then given as
\begin{equation}
  \rho_{\rm rest} \sim \frac{E}{4 \pi R^2 T \gamma^2} = 4 \times 10^5
  E_{56} \gamma_{300}^{-6} T_{10}^{-3} \ \rm erg \ cm^{-3} \ .
\end{equation}
We assume that there is a magnetic field which is nearly in equipartition
with this energy density, i.e., 
\begin{equation}
B = 3.2 \times 10^3 \xi_B^{1/2}
E_{56}^{1/2} \gamma_{300}^{-3} T_{10}^{-3/2} \ \rm Gauss \ , 
\end{equation}
where 
$\xi_B$ is a parameter of order unity for degree of equipartition between
magnetic fields and shock heated matter.

The most efficient energy loss process of protons or electrons would be 
synchrotron radiation. The photon energy of proton synchrotron in observer's
frame is given by $\varepsilon_\gamma = \varepsilon_p^2 e B/(m_p^3 \gamma)$,
where $\varepsilon_p$ is the proton energy in the observer's frame.
Therefore, if the observed 10 TeV gamma-rays are proton synchrotron,
the energy of parent protons must be
\begin{equation}
\varepsilon_p = 3.6 \times 10^{20} \left( \frac{\varepsilon_\gamma}{\rm
10 TeV} \right)^{1/2} \gamma_{300}^2 \xi_B^{-1/4} E_{56}^{-1/4} T_{10}^{3/4}
\ \rm eV \ ,
\label{eq:p-sync}
\end{equation}
which is just the energy scale of the UHECRs observed on the Earth. It
has been known that physical quantities of GRBs allow protons to be
accelerated to such high energy when $E \sim 10^{51}$ erg (Waxman
1995; Vietri 1995). We are now considering a much more energetic case
and we have to check whether protons are accelerated to $\sim 10^{20}$ eV
in the present model.  The acceleration time of Fermi acceleration is
roughly given by the gyroperiod, 
$2 \pi \eta r_L$, where $r_L = m_p \gamma_p /(eB)$ is the
Larmor radius of protons, $\gamma_p = \varepsilon_p/(m_p \gamma)$ the
Lorentz factor of protons in the shock restframe, and $\eta$ a
parameter of order unity (e.g., de Jager et al. 1996). 
The acceleration limit is given by $2 \pi \eta r_L
= R/\gamma$, where $R/\gamma$ is the expansion time measured in the
shock restframe.  (The shell thickness measured in the shell frame is
also given by $\sim T\gamma \sim R/\gamma$, and the confinement
condition is also satisfied automatically.)  Hence, the maximum proton
energy is given by 
\begin{equation}
\varepsilon_p = 4.1 \times 10^{21} \eta^{-1}
\gamma_{300}^{-1} \xi_B^{1/2} E_{56}^{1/2} T_{10}^{-1/2} \ \rm  eV \ .  
\end{equation}
Proton
acceleration is also limited by synchrotron cooling, i.e., $2 \pi \eta r_L =
t_{\rm cool}$, where $t_{\rm cool} = 6 \pi m_p^3/(\sigma_T m_e^2 B^2
\gamma_p)$ is the synchrotron cooling time at the shock rest.  This
suggests that the maximum energy achieved by proton acceleration is
\begin{equation}
\varepsilon_p = 4.4 \times 10^{20} \eta^{-1/2} \xi_B^{-1/4}
E_{56}^{-1/4} \gamma_{300}^{5/2} T_{10}^{3/4} \ \rm eV \ . 
\end{equation}
This energy is higher than
that corresponding to 10 TeV synchrotron photons (see eq. \ref{eq:p-sync}),
and hence proton
synchrotron radiation extends to $\sim$ 10 TeV range. In fact, the
maximum photon energy of synchrotron emission is determined only by
fundamental constants independently of $B$, if the maximum particle
energy is constrained by synchrotron cooling. For electron
synchrotron, the maximum photon energy is given by $\sim 25 \eta^{-1}
\gamma (\sin\theta)^{-1}$ MeV (de Jager et al.  1996), and that for
proton synchrotron is higher than this by a factor of $(m_p/m_e)$,
where $\theta$ is the pitch angle of particles along the direction of
magnetic fields. By using $\langle \sin \theta \rangle = \pi/4 = 0.785$ for
isotropic particle distribution, the maximum photon energy of proton
synchrotron is $\sim 18 \eta^{-1} \gamma_{300}$ TeV. Since this maximum
energy is quite model-independent and $\eta$ is likely larger than 1
(de Jager et al. 1996), we can infer the Lorentz factor of GRBs
from the observed maximum energy of VHE gamma-rays. The energy threshold
(50\% trigger efficiency) of the HEGRA AIROBICC array is 16 (25) TeV
(Padilla et al. 1998), and hence we obtain a lower limit of $\gamma
\gtilde 300 \eta$ and likely $\gtilde 500 \eta$. 

It should be
noted that the maximum proton energy is determined by the synchrotron
cooling constraint rather than the expansion time.  This suggests that
protons around the maximum energy will effectively lose their energy
in GRBs and hence significant fraction of energy carried by protons
will be converted into 10 TeV photons. Such VHE emission would be much
more energetic than keV--MeV emission by electrons, giving a natural
explanation for the VHE bursts observed by the two experiments.  
In fact, the
cooling time of protons for an observer is given by
\begin{eqnarray}
  t_{\rm cool, obs} = \frac{t_{\rm cool}}{2\gamma} = 0.62 \left(
    \frac{\varepsilon_\gamma}{\rm 10 TeV} \right)^{-1/2} \nonumber \\ \times
  \gamma_{300}^4 \xi_B^{-3/4} E_{56}^{-3/4} T_{10}^{9/4} \ \rm sec \ ,
\end{eqnarray}
which is sufficiently short to explain the time structure of the
observed TeV bursts with the time scale of $\sim$ 10 sec.

Next we consider whether electrons in this model can explain the
ordinary keV--MeV GRBs. The photon energy of electron
synchrotron in observer's frame is given by
\begin{equation}
  \varepsilon_\gamma = \frac{\gamma \gamma_e^2 e B}{m_e} = 1.0 \left(
    \frac{\gamma_e}{\gamma} \right)^2 \xi_B^{1/2} E_{56}^{1/2}
  T_{10}^{-3/2} \ \rm keV \ ,
\end{equation}
where $\gamma_e$ is the electron Lorentz factor measured in the shock
frame.  If the energy transfer from protons to electrons is
inefficient, which is the case we are now considering, the minimum
Lorentz factor of electrons is a direct consequence of initial bulk
motion and hence $\gamma_e \sim \gamma$. Observationally the lower
bound of energy range of GRBs is a few keV, and our model naturally
gives this lower bound.  Electrons are also accelerated to higher
energies and such electrons explain the higher energy gamma-rays
extending up to MeV--GeV.  As is done for protons above, we can
estimate the maximum energy of electrons assuming that the
acceleration time is $2 \pi \eta r_L$. The maximum electron 
energy is determined by
the synchrotron cooling constraint and hence the maximum 
photon energy of synchrotron is given by the formula: 
\begin{eqnarray}
\varepsilon_{\gamma, \max} 
&=& 25 \ \eta^{-1} \gamma \ (\sin\theta)^{-1} \ \rm MeV \\
&=& 9.5 \ \eta^{-1} \gamma_{300} \ \rm GeV \ . 
\end{eqnarray}
This is consistent with the observation of some bright GRBs extending to the
GeV range (Hurley 1996). [Proton synchrotron may also contribute
(Totani 1998a) to the GeV range gamma-ray emission, especially for the
long duration GeV emission observed for GRB940217 (Hurley et
al. 1994).]  The cooling time of electrons for an observer is written as:
\begin{eqnarray}
t_{\rm cool, obs} = \frac{t_{\rm cool}}{2\gamma} = 
1.36 \times 10^{-1} \left( \frac{\varepsilon_\gamma}{\rm 10 keV}\right)^{-1/2}
\nonumber \\ \times
   \gamma_{300}^4 \xi_B^{-3/4} E_{56}^{-3/4} T_{10}^{9/4} \ \rm msec \ ,
\end{eqnarray}
which is near the shortest time variability observed in GRBs ($\sim$ msec). 
Hence we can interpret the msec time scale of variability as 
synchrotron cooling time of electrons. This hypothesis predicts
that the time scale of variability changes with photon energy as 
$\propto \varepsilon_\gamma^{-1/2}$, which can be tested by existing data.

We have shown that the recent observations of 10 TeV gamma-rays from some
GRBs can be explained quite well if protons carry much larger energy than
electrons and protons are accelerated up to $\sim 10^{20-21}$ eV. Furthermore, 
electron synchrotron radiation in our model naturally gives the correct 
energy range and time variability of the ordinary keV--MeV gamma-rays
of GRBs. It seems quite difficult to explain such
strong emission of 10 TeV gamma-rays by other radiation processes
and hence the observations of the two 
independent experimental groups suggest that
protons in GRBs are actually accelerated up to the energy scale of the
UHECRs observed on the Earth. The flux
of UHECRs is difficult to estimate theoretically because the escape fraction
of such protons from GRBs is poorly known and the lifetime of UHECRs
in the intergalactic space drastically changes 
around $10^{20}$ eV. This is known as the Greisen-Zatsepin-Kuzmin
cutoff (Greisen 1966; Zatsepin \& Kuzmin 1966), which is due to the photopion
production by the interaction between UHECRs and the 2.7K
cosmic microwave background radiation.
The lifetime is about $\tau_{\rm CR} \sim 10^8$ yrs
for protons with energy of $\sim 2 \times 10^{20}$ eV (Berezinski\u\i \ 
et al. 1990), corresponding to a horizon for UHECRs at $\sim 30$ Mpc.
However, because of the
large total energy of $\sim 10^{56}b^{-1}$ erg for one GRB, where
$b = 4\pi/\Delta\Omega$ is the beaming factor, it can be concluded
that the production rate of UHECRs in GRBs
is sufficiently large to explain the observed UHECR flux above $10^{20}$ eV.
In fact, we can calculate the energy production of UHECRs per one GRB
required to explain the observed UHECR flux as
\begin{eqnarray}
E_{\rm CR} = 2 \times 10^{54} b^{-1} 
\left( \frac{\rho_{\rm CR}}{7 \times 10^{-21}
\rm \ erg \ cm^{-3}} \right) \nonumber \\ \times
\left( \frac{\tau_{\rm CR}}{\rm 10^8 \ yr} 
\right)^{-1} \left( \frac{R_{\rm GRB}}{10^{-9} b \rm \ yr^{-1} Mpc^{-3}} 
\right)^{-1} \ \rm erg \ ,
\end{eqnarray}
where $\rho_{\rm CR}$ is the energy density of UHECRs deduced from
the observed UHECR flux (Yoshida and Dai 1998), and
$R_{\rm GRB}$ the present-day rate density of GRBs (Totani 1997; 1998b).
If the total energy liberated by one GRB is $\sim 10^{56}b^{-1}$ erg,
UHECR production with $E_{\rm CR} \sim 10^{54}b^{-1}$ erg would be quite
reasonable and hence we conclude that the 10 TeV gamma-rays observed by
the Tibet and HEGRA groups give a strong support for the hypothesis that
UHECRs are produced by GRBs. 

The author would like to thank an anonymous referee
for useful and constructive comments.
He has been supported by the Research Fellowships of the Japan
Society for the Promotion of Science for Young Scientists, and
the Grant-in-Aid for the
Scientific Research Fund (No. 3730) of the Ministry of Education, Science,
and Culture of Japan.

%\begin{thebibliography}{DUM}

\end{document}